\documentclass{jfm}
\usepackage{graphicx}

\usepackage[amssymb]{SIunits}
\usepackage{amsmath}
\usepackage[usenames,dvipsnames]{color}

\usepackage{xcolor}

\newcommand{\eref}[1]{equation~\eqref{#1}}

\newcommand{\fref}[1]{figure~\ref{#1}}

\makeatletter
\def\mathcolor#1#{\@mathcolor{#1}}
\def\@mathcolor#1#2#3{%
  \protect\leavevmode
  \begingroup
    \color#1{#2}#3%
  \endgroup
}
\makeatother

\shorttitle{Turbulence strength in ultimate Taylor-Couette turbulence}
\shortauthor{R. Ezeta et al}

\title{Turbulence strength in ultimate Taylor-Couette turbulence}

\author{
  Rodrigo Ezeta\aff{1},
  Sander G. Huisman\aff{1},
 Chao Sun\aff{3,1}
   \corresp{\email{chaosun@tsinghua.edu.cn}}
  ,
 \and Detlef Lohse\aff{1,4}
 }

\affiliation{
\aff{1}Physics of Fluids Group, MESA$^{+}$ Institute  and J.M. Burgers Centre for Fluid Dynamics, University of Twente, P.O. Box 217, 7500AE Enschede, The Netherlands
\aff{3}Center for Combustion Energy and Department of Thermal Engineering, Tsinghua University, Beijing 100084, China
\aff{4}Max Planck Institute for Dynamics and Self-Organisation, 37077 G\"{o}ttingen, Germany
}

\begin{document}

\maketitle

\begin{abstract}
We provide experimental measurements for the effective scaling of the Taylor-Reynolds number within the bulk $\text{Re}_{\lambda,\text{bulk}}$, based on local flow quantities as a function of the driving strength (expressed as the Taylor number Ta), in the ultimate regime of Taylor-Couette flow. The data are obtained through flow velocity field measurements using Particle Image Velocimetry (PIV). We estimate the value of the local dissipation rate $\epsilon(r)$ using the scaling of the second order velocity structure functions in the longitudinal and transverse direction within the inertial range---without invoking Taylor's hypothesis. We find an effective scaling of $\epsilon_{\text{bulk}} /(\nu^{3}d^{-4})\sim \text{Ta}^{1.40}$, (corresponding to $\text{Nu}_{\omega,\text{bulk}} \sim \text{Ta}^{0.40}$ for the dimensionless local angular velocity transfer), which is nearly the same as for the global energy dissipation rate obtained from both torque measurements ($\text{Nu}_{\omega} \sim \text{Ta}^{0.40}$) and Direct Numerical Simulations ($\text{Nu}_{\omega} \sim \text{Ta}^{0.38}$). The resulting Kolmogorov length scale is then found to scale as $\eta_{\text{bulk}}/d \sim \text{Ta}^{-0.35}$ and the turbulence intensity as $I_{\theta,\text{bulk}} \sim \text{Ta}^{-0.061}$. With both the local dissipation rate and the local fluctuations available we finally find that the Taylor-Reynolds number effectively scales as Re$_{\lambda,\text{bulk}}\sim \text{Ta}^{0.18}$ in the present parameter regime of $4.0 \times 10^8 < \text{Ta} < 9.0 \times 10^{10}$.

\end{abstract}


\section{Introduction}
\label{sec:introduction}

Taylor-Couette (TC) flow, the flow between two coaxial co- or counter-rotating cylinders, is one of the idealized systems in which turbulent flows can be paradigmatically studied due to its simple geometry and its resulting accessibility through experiments, numerics, and theory. In its rich and vast parameter space, various different flow structures can be observed \citep{GITaylor1923,Chandrasekhar1961,Andereck1986,VanGils2011,Huisman2014,Rodolfo2014,veen2016a}. For recent reviews, we refer the reader to \citet{Fardin2014} for the low $\text{Ta}$ range and \citet{Grossmann2016} for large $\text{Ta}$.

The driving strength of the system is expressed through the Taylor number defined as

\begin{equation}
\nonumber
\label{eq:ta} 
\text{Ta}= \frac 14 \sigma_{\text{TC}} d^2(r_i+r_o)^2(\omega_i-\omega_o)^2 / \nu^2,
\end{equation}

\noindent where $r_{i,o}$ are the inner and outer radii, $d=r_o-r_i$ the gap width, $\omega_{i,o}$ the angular velocities of the inner and outer cylinders, $\nu$ the kinematic viscosity of the fluid, $\sigma_{\text{TC}}=(1+\rho)^4/(4\rho)^2\approx1.06$ a pseudo-Prandtl number employing the analogy with Rayleigh-B\'{e}nard (RB) flow \citep{Eckhardt2007}, and $\rho=r_i/r_o$ the radius ratio. The response of the system is generally described by the two response parameters $\text{Nu}_\omega$ and $\text{Re}_w$. The first is the Nusselt number $\text{Nu}_{\omega}=J_{\omega}/J_{\omega,\text{lam}}$, with the angular velocity transfer $J_{\omega}=r^3\langle (u_r\omega-\nu\partial_r\omega)\rangle_{A,t}$, where $\langle \rangle_{A,t}$ denotes averaging over a cylindrical surfaces of constant radius and over time. $\omega=u_\theta/r$ is the angular velocity and $J_{\omega,\text{lam}}=2\nu(r_ir_o)^2(\omega_i-\omega_o)/(r_o^2-r_i^2)$ is the angular velocity transfer from the inner to the outer cylinder for laminar flow. $\text{Nu}_\omega$ describes the flux of angular velocity in the system, and is directly linked to the torque through the Navier-Stokes equations. The second response parameter of the flow is the so-called wind Reynolds number $\text{Re}_w=\sigma_{\text{bulk}}(u_r)d/\nu$, where $\sigma_{\text{bulk}}(u_r)$ is the standard deviation of the radial component of the velocity inside the bulk. $\text{Re}_w$ quantifies the strength of the secondary flows.
In the ultimate regime of turbulence, where both the boundary layers (BL) and the bulk are turbulent (Ta $\geq 3\times 10^8$), it was experimentally found that $\text{Nu}_\omega \sim \text{Ta}^{0.40}$, in the Taylor number regime of $10^9$ to $10^{13}$, independent of the rotation ratio $a=-\omega_o/ \omega_i$ and radius ratio $\rho$ \citep{VanGils2011,Huisman2014,Paoletti2011,Rodolfo2014}. This scaling has been identified, using the analogy with RB flow, with the ultimate scaling regime $\text{Nu}_\omega\sim \text{Ta}^{1/2} \mathcal{L}(\text{Ta})$, where the log-corrections $\mathcal{L}(\text{Ta})$ are due to the presence of the BLs. \citep{Grossmann2011}. The wind Reynolds number $\text{Re}_w$ was found experimentally to scale as $\text{Re}_w\sim \text{Ta}^{0.495}$ within the bulk flow \citep{Huisman2012}; very close to the $1/2$ exponent that was theoretically predicted by \citet{Grossmann2011}. Here, remarkably, the log-corrections cancel out.

In this study we characterize the local response of the flow with an alternate response parameter based on the standard deviation of the azimuthal velocity $\sigma(u_\theta)$ and the microscales of the turbulence, i.e. the Taylor-Reynolds number which is defined as $\text{Re}_\lambda=u'\lambda/\nu$, where $u'$ is the rms of the velocity fluctuations and $\lambda$ is the Taylor micro-scale. 

$\text{Re}_\lambda$ is often used in the literature to quantify the level of turbulence in a given flow, ideally for homogeneous and isotropic turbulence (HIT), where it should be calculated from the full 3D velocity field. In experiments however, the entire flow field is generally not accessible. Assuming isotropy (which is most of the time not strictly fulfilled), the dissipation rate $\epsilon$ (in Cartesian coordinates) can be reduced to $\epsilon=15\nu \langle (\partial u/ \partial x)^2 \rangle_{t}$, where $u$ is the component of the velocity in the streamline direction $x$. In this way, the Taylor micro-scale is then redefined as $\lambda=\langle u^2 \rangle  / \langle (\partial u / \partial x)^2 \rangle  $. Examples where this procedure has been followed in spite of the lack for perfect isotropy include turbulent RB flow \citep{zhou2008}, the flow between counter-rotating disks \citep{Voth2002}, von K\'{a}rm\'{a}n flow  \citep{zimmermann2010}, or channel flow \citep{martinez2012a}. In all cases the isotropic form of $\text{Re}_{\lambda}$ is still chosen as a robust way to quantify the strength of the turbulence. It is in this spirit that we aim to calculate $\text{Re}_{\lambda}$ in turbulent Taylor-Couette flow, albeit in a region sufficiently far away from the BLs (bulk). Such a calculation allows for a quantitative comparison between the turbulence generated in TC flow and the one produced by other canonical flows, i.e. pipe, channel, RB, von K\'{a}rm\'{a}n flow, etc. Following this route, we define the bulk Taylor-Reynolds number for TC flow as

\begin{align}
\text{Re}_{\lambda,\text{bulk}} &\equiv (\sigma_{\text{bulk}}(u_\theta))^2\left( \frac{15}{\nu \epsilon_{\text{bulk}}} \right)^{1/2} \label{eq:tare},\\
\sigma_{\text{bulk}}(u_\theta)  &\equiv \left \langle \sigma_{\theta,t} \left( u_\theta(r,\theta,t) \right) \right \rangle_{r_{\text{bulk}}} \label{eq:sigma1},\\
\epsilon_{\text{bulk}} &\equiv \left \langle \epsilon(r,\theta,t) \right \rangle_{\theta,t,r_{\text{bulk}}} \label{eq:sigma2},
\end{align}

\noindent where $\sigma_{\theta,t} \left( u_\theta(r,\theta,t) \right)$ is the standard deviation of the azimuthal velocity in the azimuthal direction and over time. $\sigma_{\text{bulk}}(u_\theta)$ is then the average of the azimuthal velocity fluctuations profile over the bulk and $\epsilon_{\text{bulk}}$ the bulk-averaged dissipation rate. Note that the subscript $r_{\text{bulk}}$ means that we average in the radial direction but only for $0.35 < (r - r_i)/d < 0.65$, i.e.~the middle 30\% of the gap (see also \S\ref{sec:bulkregion}).
   
Multiple prior estimates of $\text{Re}_\lambda$ in TC flow can be found in the literature:  \citet{Huisman2013} calculated it using a combination of the  local velocity fluctuations and the global energy dissipation rate $\epsilon_{\text{global}}$, where the latter is obtained from torque measurements denoted by $\tau$ through $\epsilon=\tau \omega_i/m$, where $m$ is the total mass. \citet{Lewis1999}, however, estimated $\text{Re}_\lambda$ at midgap ($\tilde{r}=(r-r_i)/d=0.5$) with the local velocity fluctuations and a local dissipation rate estimated indirectly through the velocity spectrum $E(k)$ in wave number space $k$, i.e. $\epsilon=15 \nu \int k^2 E(k)dk$. In this calculation, Taylor's frozen flow hypothesis was used to get the $\theta$-dependence for the azimuthal velocity $u_\theta$, i.e. $u(\theta+d\theta,t)=u(\theta,t-rd\theta/U)$, where $U$ is the mean azimuthal velocity. To the best of our knowledge, however, a truly bulk-averaged calculation of $\text{Re}_{\lambda,\text{bulk}}$ (based on local quantities) has hitherto never been reported in the literature. Of particular interest is how this quantity scales with $\text{Ta}$ in the ultimate regime, and how this scaling is connected to that of $\text{Nu}_\omega$ and $\text{Re}_w$. 

As TC flow is a closed flow system, the global energy dissipation rate $\epsilon_{\text{global}}$ is connected to both the driving strength Ta and $\text{Nu}_{\omega}$ by \citep{Eckhardt2007}

\begin{equation}
\label{eq:epsilon_global}
\tilde{\epsilon}_\text{global}= \frac{d^4}{\nu^3} \epsilon_\text{global}   = \sigma_{\text{TC}}^{-2}  \text{Nu}_\omega \text{Ta}.
\end{equation}

In the ultimate regime this implies an effective scaling of the global energy dissipation rate $\tilde{\epsilon}_\text{global}\sim \text{Ta}^{1.40}$. A calculation of $\text{Re}_\lambda$ in the bulk does not require the global energy dissipation rate $\tilde{\epsilon}_\text{global}$, but the bulk-averaged energy dissipation rate, $\epsilon_{\text{bulk}}$ in combination with the bulk averaged velocity fluctuations $\sigma_{\text{bulk}}(u_\theta)$, see \eref{eq:sigma1}. In general, velocimetry techniques like Particle Image Velocimetry (PIV) can provide $\sigma_{\text{bulk}}(u_\theta)$ directly, thus the challenge of the calculation is to correctly estimate $\epsilon_{\text{bulk}}$. While the global energy dissipation rate $\epsilon_{\text{global}}$ (\eref{eq:epsilon_global}) can be obtained from torque measurements, an estimate of $\epsilon_{\text{bulk}}$ requires the knowledge of the local dissipation rate $\epsilon(r,\theta,t)$ as it is shown in \eref{eq:sigma2}.  For fixed height along the cylinders, the dissipation rate profile $\epsilon(r)=\langle \epsilon(r,\theta,t) \rangle_{\theta,t}$ is connected to the global energy dissipation rate through $\epsilon_\text{global}=(\pi ( r_o^2-r_i^2))^{-1} \int_{r_i}^{r_o} \epsilon (r) 2\pi r dr$. We note that due to the non-trivial interplay between bulk and turbulent BLs in the ultimate regime, it is not known a priori that $\epsilon_{\text{bulk}}$ and $\epsilon_\text{global}$ will scale in the same way: local measurements are needed to confirm this assumption.

The energy dissipation rate $\epsilon$ is key for Kolmogorov's scaling prediction of the velocity structure functions (SFs) in HIT, namely $D_{LL}(s)=C_2 (\epsilon s)^{2/3}$ for the second order longitudinal structure function and $D_{NN}(s)=C_2 (4/3)(\epsilon s)^{2/3}$ for the second order transverse structure function within the inertial range, neglecting intermittency corrections \citep{Pope2000,Frisch1995}. The Kolmogorov constant was measured to be $C_2\approx 2.0$ and is believed to be universal \citep{Sreenivasan1995}. The exponents for the scaling of the $p$-th order SFs ($\zeta_p^{\star}$) have been measured and found to differ from Kolmogorov's original prediction $p/3$: the difference between them are attributed to the intermittency of the flow \citep{Benzi1993,Z.She1994,Lewis1999,Huisman2013}. However, second order SFs along with the classical Kolmogorov scaling $\zeta_2=2/3$ have been successfully used to estimate $\epsilon$ in fully developed turbulence \citep{Voth2002,Blum2010,zimmermann2010,Chien2013}. One can then expect only a moderate underestimation of $\epsilon$ since the intermittency correction to the exponent of the second order SFs is small $\zeta_2^{\star}-2/3 \approx 0.03$, where $\zeta_2^{\star}$ is the measured exponent of the second order SFs in TC flow using extended self-similarity (ESS) \citep{Lewis1999,Huisman2013}.

In this paper we make use of local flow measurements using planar Particle Image Velocimetry (PIV) to find $\sigma_{\text{bulk}}(u_\theta)$ and using the scaling  of the second order ($p=2$) SFs we estimate $\epsilon_{\text{bulk}}$. The advantage of PIV over other flow measuring technique such as Laser- Doppler or Hot-wire anemometry is the possibility to access the whole velocity field at the same time in the $r-\theta$ plane, i.e. $\vec{u}=u_r(r,\theta,t)\hat{e_r}+u_\theta(r,\theta,t)\hat{e_\theta}$,  from which we can obtain directly the  $\theta$-dependence of the velocities. Unlike in the calculation of \citet{Lewis1999} and \cite{Huisman2013}, in this work, we do not need to invoke Taylor's hypothesis in the calculation of $\text{Re}_{\lambda,\text{bulk}}$. We only explore the case of inner cylinder rotation ($a=0$), where there is virtually no stable structures (Taylor rolls) left when the driving strength is sufficiently large ($\text{Ta}\geq 10^8$) \citep{Huisman2014}. In this way, the calculation is independent of the axial height $z$ and thus there is no need for an axial average \citep{vangils2012}.

\section{Experimental apparatus}
\label{sec:setup}

\begin{figure}
\begin{center}
\includegraphics[scale=0.4]{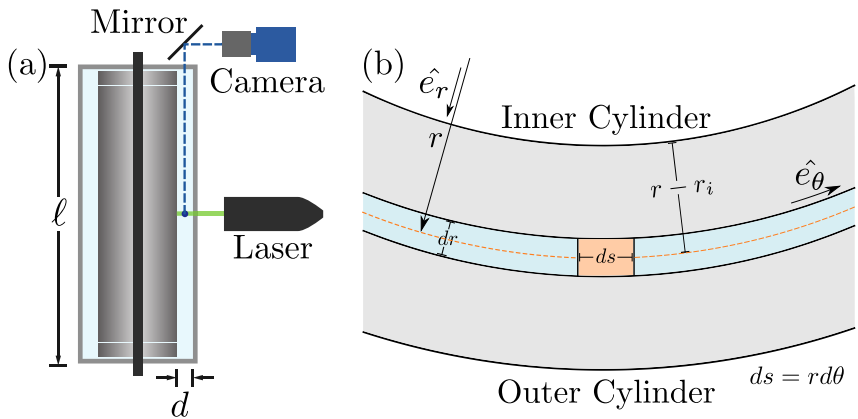}
\caption{(color online). (a) Vertical cross section of the experimental setup. (b) A sketch of the binning process on the $r-\theta$ plane for the calculation of the SFs. Here we show an exaggeration of how the velocity fields are binned in both the radial and azimuthal directions. $\hat{e_r}$  and $\hat{e_\theta}$ are the unit vectors in polar coordinates. The orange dashed line represents the streamline direction $s$ for a fixed radius.}
\label{fig:figure1}
\end{center}
\end{figure}

The PIV experiments were performed in the Taylor-Couette apparatus as described in \citet{Huisman2015}. This facility provides an optimal environment for PIV experiments in TC flow, due to its transparent outer cylinder and top plate. The radii of the setup are $r_i=\unit{75}{\milli \meter}$ and $r_o=\unit{105}{\milli \meter}$, and thus $\rho=r_i/r_o=0.714$, which is very close to $\rho=0.724$ and $\rho=0.716$ from \citet{Lewis1999} and \citet{Huisman2013}, respectively. The height  $\ell$ equals $\unit{549}{\milli \meter}$,  resulting in an aspect ratio $\Gamma=\ell/d=18.3$. The excellent temperature control of the setup allows us to perform all the experiments at a constant temperature of $\unit{26.0}{\celsius}$ with a standard deviation of $\unit{15}{\milli \kelvin}$. The measurements are done at midheight $z=\ell/2$ in the $r-\theta$ plane. The flow is seeded with fluorescent polyamide particles with diameters up to 20 $\mu$m and with an average particle density of $\approx 0.01 \ \text{particles/px}$. The laser sheet we use for illumination is provided by a pulsed laser (Quantel Evergreen 145 laser, 532 nm) and has a thickness of $\approx 2.0$ mm. The measurements are recorded using a high-resolution camera  at a framerate of $f=1$ Hz. The camera we use is an Imager sCMOS $(2560\  \text{px} \times 2160 \ \text{px})$ 16 bit with a Carl Zeiss Milvus 2.0/100. The camera is operated in double frame mode which leads to an inter-frame-time  $\Delta t\ll 1/f$. In \fref{fig:figure1}a  a schematic of the experimental setup is shown. In order to obtain a large amount of statistics, we capture 1500 fields for each of the 12 different Taylor numbers explored. The velocity fields are calculated using a ``multi-pass" method with a starting window size of $\unit{64}{px}\times\unit{64}{px}$ to a final size of  $\unit{24}{px}\times\unit{24}{px}$ with $50\%$ overlap. This allows us to obtain  a resolution of $d x=0.01d$. When using the local Kolmogorov length scale in the flow (see \S\ref{sec:epsilon}), we find that $d x / \eta_{\text{bulk}}$ ranges from $\approx1.6$ ($\text{Ta}=4.0\times 10^{8}$) to $\approx 10$ ($\text{Ta}=9.0\times 10^{10}$).

\section{Results}

\subsection{Identifying the bulk region}
\label{sec:bulkregion}

\begin{figure}
\begin{center}
\includegraphics[scale=0.35]{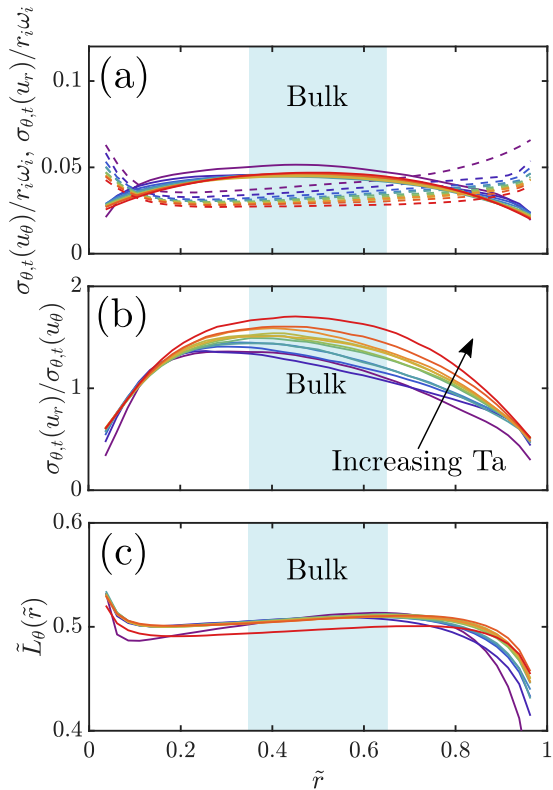}
\caption{(color online). (a) Normalized velocity fluctuations profiles for various Ta: (dashed lines) azimuthal, (solid lines) radial. (b) The profiles of the velocity fluctuations ratio (radial/azimuthal) for various Ta. (c) Normalized specific angular momentum profile for various Ta. In all figures, the bulk region $\tilde{r} \in [0.35,0.65]$ is highlighted as the blue region. The different colors represent different Ta as described in \fref{fig:figure3}. }
\label{fig:figure2}
\end{center}
\end{figure}

The profiles of the velocity fluctuations for both components of the velocity as a function of Ta are shown in \fref{fig:figure2}a. The distance from the inner cylinder is represented by the normalized radius  $\tilde{r}=(r-r_i)/d$. When normalized with the velocity of the inner cylinder $r_i \omega_i$, both profiles collapse for all Ta numbers in most of the gap width around the value of 0.03. Only very close the inner and outer cylinder, the fluctuations increase (decrease) for the azimuthal (radial) component. In our calculation of $\text{Re}_{\lambda,\text{bulk}}$ (\eref{eq:tare}), we use $\sigma_{\text{bulk}}(u_\theta)$ as our velocity scale as $u_\theta$ is the primary flow direction. Here, we are essentially assuming that the radial and axial velocity fluctuations, on average, have the same order of magnitude, i.e. $\sigma_{\text{bulk}}(u_\theta)\approx \sigma_{\text{bulk}}(u_r)$ (the result is z-independent). In order to give an impression of how valid this assumption is, in \fref{fig:figure2}b we show the ratio of the velocity fluctuations throughout the gap. We notice that within the bulk region, the ratio is between $1.0$ and $1.6$ for all analyzed Ta numbers; consistent with what one would expect for reasonably isotropic flows. Surprisingly, the ratio within the bulk increasingly deviates from unity as the driving is increased. The same observation is also observed in turbulent TC-flow ($\text{Ta} \in[5.8\times 10^7, 6.2\times 10^9]$) for a wider gap $\eta=0.5$, where also the ratio within the bulk increasingly deviates from unity with increasing Ta. In that case however, it seems to reach a value of  $\approx 1.8$ for the largest Ta \citep{vdveen2016}. Since the same observation is found in two different studies (with two different experimental setups), we believe this is a feature of TC-flow; however, a more rigorous theoretical explanation has yet to be provided. Another interesting feature of the profiles in \fref{fig:figure2}b is that they become flatter as the turbulence level is increased, reflecting an increase in spatial homogeneity. Note that these results do not suggest readily that the flow is in a HIT state. What this merely shows is that there is a special region (bulk) where the flow becomes more homogeneous as compared to regions close to the solid boundaries and it is reasonably isotropic. This justifies that our calculation is based on an isotropic form of $\text{Re}_\lambda$ as was also used in other studies \citep{Lewis1999,Voth2002,zhou2008,zimmermann2010,martinez2012a}.

Next, we define the bulk region as $r_\text{bulk}\equiv r-r_i  \in [0.35d,0.65d]$, wherein the magnitude of the velocity fluctuations for both $u_r$ and $u_\theta$ are roughly constant. This definition of the bulk was previously used by \citet{Huisman2012} who measured the scaling of $\text{Re}_w$ in the ultimate regime. The same definition is also consistent with other studies \citep{Smith1982,Lewis1999}, where the bulk region is identified as the $r$ domain wherein the normalized specific angular momentum remains constant ($\tilde{L}_{\theta}=r \langle u_\theta\rangle_{\theta,t}/(r_i^2 \omega_i) \approx 0.5$) for all Ta. In \fref{fig:figure2}c we show $\tilde{L}_{\theta}(r)$ and we find a good collapse of the profiles within our definition of the bulk. Here, it is seen that the value of $\tilde{L}_\theta$ is indeed around 0.5 within the bulk.

\subsection{Structure functions and energy dissipation rate profiles }
\label{sec:sfs}
Having defined the bulk region, we bin the velocity data in the azimuthal (streamwise) direction with a bin width $d\theta=0.2^\circ$ for every $r$ and Ta. Now we calculate the second order structure functions in both longitudinal (LL) and transverse (NN) directions for every radial bin,

\begin{eqnarray}
\label{eq:sf}
\delta_{LL}(r,s)&=& \langle  (u_{\theta}(r,\theta +s/r,t)-u_{\theta}(r,\theta,t))^2 \rangle_{\theta,t}  \\
\delta_{NN}(r,s)&=& \langle (u_{r}(r,\theta +s/r,t)-u_{r}(r,\theta,t))^2 \rangle_{\theta,t}, 
\end{eqnarray}

\begin{figure}
\begin{center}
\includegraphics[scale=0.35]{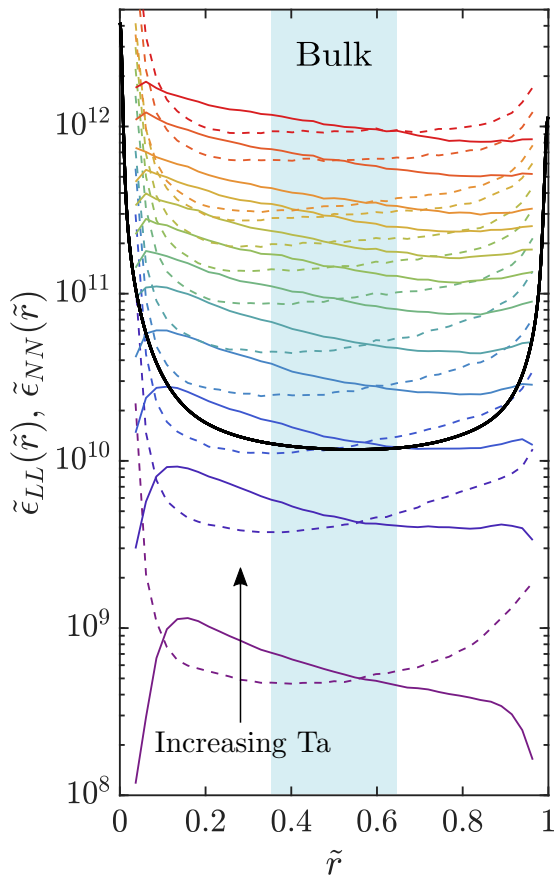}
\caption{(color online). Dimensionless energy dissipation rate profile $\tilde{\epsilon}(r) = \epsilon(r)/(d^{-4}\nu^3)$ for various Ta: (dashed lines) longitudinal  direction $\tilde{\epsilon}_{LL}(\tilde{r})$, (solid lines) transversal direction $\tilde{\epsilon}_{NN}(\tilde{r})$. Ta is increasing from bottom to top, the lines correspond to the following Ta numbers: $\text{Ta}=4.0\times 10^{8}$, $1.6\times 10^{9}$, $3.6\times 10^9$, $6.4\times 10^9$, $1.0\times 10^{10}$, $1.4 \times 10 ^{10}$, $2.0 \times 10^{10}$, $2.6\times 10^{10}$, $3.2\times 10^{10}$, $4.0 \times 10^{10}$, $5.7\times 10^{10}$, $9.0\times 10^{10}$. For every Ta, both $\epsilon$-profiles cross within the bulk region ($\tilde{r}\in[0.35,0.65] $) which is highlighted in blue. The black solid line is the total energy dissipation rate obtained from DNS for $\text{Ta}=2.15\times 10^9$ \citep{zhu2017}.}
\label{fig:figure3}
\end{center}
\end{figure}

\noindent where $s$ is the distance along the streamwise direction. Since $s=r \theta$, the azimuthal binning guarantees a constant spatial resolution $ds=r d \theta$ along the direction of $s$, when the radial variable $r$ is fixed (see the sketch in \fref{fig:figure1}b). The choice of $ds$ is limited by the resolution of the PIV experiments $dx$ and it is chosen such as to not filter out any intermittent fluctuations in the flow.

The energy dissipation rate profiles for both directions are calculated as follows. For fixed $r$ and Ta, $\epsilon_{LL}$ is chosen as the maximum of $s^{-1}(\delta_{LL}(r,s)/C_2)^{-2/3}$ such that $s$ lies inside the inertial range. In the same manner, $\epsilon_{NN}$ is taken as the maximum of $s^{-1}(\delta_{NN}(r,s)/(4C_2/3))^{-2/3}$ with the same restriction for $s$. This operation is repeated for every $r$ and Ta, leading to the dissipation rate profiles shown in \fref{fig:figure3}. In this figure, the $\epsilon$-profiles are made dimensionless as $\tilde{\epsilon}(r)=\epsilon(r)/(d^{-4}\nu^3)$. Near the solid boundaries, this figure shows that the dissipation rates (LL and NN) differ from each other: $\epsilon_{LL} $ increases while $\epsilon_{NN}$ decreases, which is consistent with the measurement of the velocity fluctuations (figures \ref{fig:figure2}ab). However, as one moves into the bulk region, the discrepancy between them decreases until eventually both dissipation rates intersect. The crossing remains within the bulk region, independent of Ta, and does not seem to occur at any particular radial position. Only in the case of HIT, the dissipation rates obtained from both SFs are exactly the same. However, as indicated in figures \ref{fig:figure2}ab, the flow tends to be more homogeneous within the bulk. We expect then that, regardless of the structure function (longitudinal or transverse) used, the energy dissipation rate obtained from either direction should, on average, be nearly the same within the bulk. In this study we will show that this is indeed the case, which means that $\epsilon_\text{bulk}$ can be obtained either from the dissipation rate in the LL direction $\epsilon_{LL}$ or from that in the NN direction $\epsilon_{NN}$. A similar approach is followed in \citet{Ni2011}, where both SFs are calculated in RB flow within the sub-Kolmogorov regime where the flow is found to be nearly homogeneous and isotropic at the center of the cell. 

In \fref{fig:figure3}, we have included the dimensionless dissipation rate $\tilde{\epsilon}_u=(d^{4}/\nu^3)\langle (\nu/2) (\partial u_i/ \partial x_j + \partial u_j/ \partial x_i)^{2} \rangle_{V,t}$ obtained from Direct Numerical Simulations (DNS) for $\rho=0.714$, $\Gamma=2$ and $\text{Ta}=2.15\times 10^{9}$ from \citet{zhu2017}. Here, the $\langle \rangle_{V,t}$ denotes the average over the entire volume and time respectively. This includes the boundary layers, that we explicitly avoid in our $r_{\text{bulk}}$ definition. When comparing the profile obtained from numerics and from our data for $\text{Ta}=3.6\times 10^{10}$ we notice that both agree rather well, thus mutually validating each other.

\begin{figure}
\begin{center}
\includegraphics[scale=0.31]{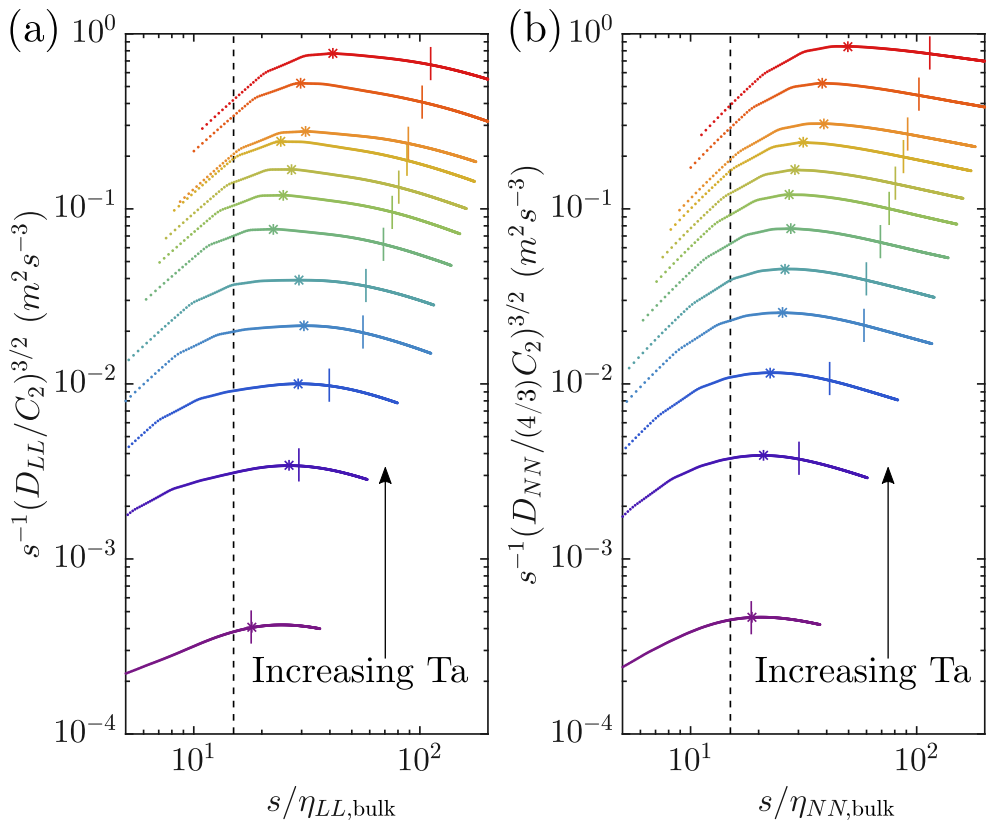}
\caption{(color online). Compensated time-bulk averaged structure functions for various Ta:  (a) longitudinal, (b) transverse. The colors represent the variation in Ta as described in \fref{fig:figure3}. In both figures, the black dashed line is 15$\eta$ while the colored short vertical lines are located at $\text{L}_{11}/\eta$ for each Ta: The inertial range is approximately bounded by these two lines. The colored stars show the maxima of each curve which correspond to $\langle \epsilon(r)\rangle_{r_\text{bulk}}$.}
\label{fig:figure4}
\end{center}
\end{figure}

By averaging the $\epsilon$-profiles in the bulk (\fref{fig:figure3}), we finally find the bulk-averaged dissipation rates $\tilde{\epsilon}_{LL,\text{bulk}}=\langle \tilde{\epsilon}_{LL}(\tilde{r})\rangle_{r_\text{bulk}} $ and $\tilde{\epsilon}_{NN,\text{bulk}}=\langle \tilde{\epsilon}_{NN}(\tilde{r})\rangle_{r_\text{bulk}} $. In order to validate the calculation, in \fref{fig:figure4} we show the bulk-averaged longitudinal $D_{LL}$ and transverse $D_{NN}$ SFs for every Ta. Here, we compensate the SFs as $s^{-1}(D_{LL}(s)/C_2)^{2/3}$ and $s^{-1}(D_{NN}(s)/(4/3)C_2)^{2/3}$ such that their units match that of the dissipation rate. The horizontal axis is normalized with the corresponding bulk-averaged Kolmogorov length scale (see \S\ref{sec:epsilon}). According to Kolmogorov's scaling, within the inertial regime ($s\in[15\eta,\text{L}_{11}]$), where $L_{11}$ is the integral length scale obtained from the azimuthal velocity, each compensated curve (fixed $r$ and Ta) should be proportional to the dissipation rate in the bulk. Here we see that our estimates for the bulk-averaged dissipation rates are located within the plateau regions, demonstrating the self-consistency of the calculation. In the same figure, the separation of length scales in the flow can also be seen. Note in particular how such separation between $\eta$ and $\text{L}_{11}$ increases with Ta. The integral length scale $\text{L}_{11}(\text{Ta})$ in \fref{fig:figure4} is calculated using the integral of the autocorrelation of the azimuthal velocity in the azimuthal direction and averaged over the bulk region.

\subsection{The dissipation rate in the bulk}
\label{sec:epsilon}

\begin{figure}
\begin{center}
\includegraphics[scale=0.4]{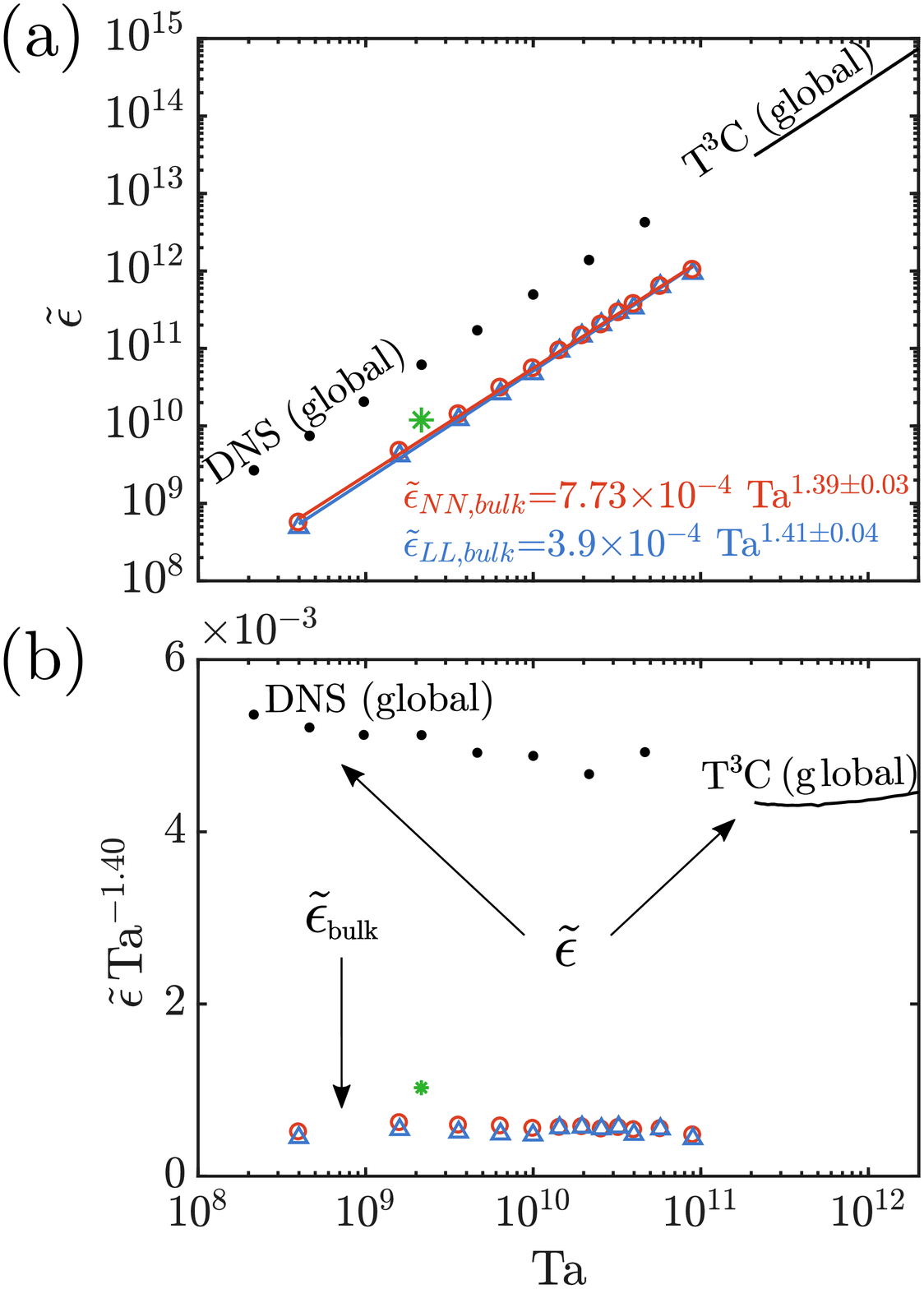}
\caption{(color online) (a) Dimensionless bulk-averaged energy dissipation rate: longitudinal $\tilde{\epsilon}_{LL,\text{bulk}}$ (blue open triangles), transverse $\tilde{\epsilon}_{NN,\text{bulk}}$ (red open circles). Dimensionless global energy dissipation rate ($\tilde{\epsilon}_{\text{global}}$): DNS \cite{Ostilla-Monico2014a} (solid black circles), torque measurements \cite{Huisman2014} (black line).  (b) Compensated plot of the bulk-averaged dissipation rate, where an effective scaling of $\tilde{\epsilon}_{\text{bulk}}\sim \tilde{\epsilon} \sim\text{Ta}^{1.40}$ is revealed for both the global and the dissipation rate in the bulk. In both figures, the green star corresponds to the bulk-average dissipation rate data of \cite{zhu2017} for $\text{Ta}=2.15\times 10^9$. }
\label{fig:figure5}
\end{center}
\end{figure}

In \fref{fig:figure5}a we show the scaling of both $\tilde{\epsilon}_{LL,\text{bulk}}$ and $\tilde{\epsilon}_{NN,\text{bulk}}$. We find that the dissipation rate extracted from both directions scale effectively as $\tilde{\epsilon}_{\text{bulk}} \sim \text{Ta}^{1.40}$, with a nearly identical prefactor. This shows that the local energy dissipation rate scales in the same way as the global energy dissipation rate $\tilde{\epsilon} \sim \text{Ta}^{1.40}$. Correspondingly, this implies that the local Nusselt number scales as $\text{Nu}_{\omega,\text{bulk}}\sim \text{Ta}^{0.40}$. In the same figure (\fref{fig:figure5}a), we include $\tilde{\epsilon}$ of \citep{Ostilla-Monico2014a}, obtained from both DNS, and \cite{Huisman2014} torque measurements from the Twente Turbulent Taylor-Couette  (T$^3$C) experiment. The compensated plot (\fref{fig:figure5}b) reveals that both the local and global energy dissipation rate scale indeed as $\text{Ta}^{1.40}$ with the ratio $ \epsilon_{\text{bulk}} / \epsilon_{\text{global}}  \approx 0.1 $. In the regime of ultimate TC turbulence, it was suggested that both turbulent BLs extend throughout the gap until they meet around $d/2$ \citep{Grossmann2011}. The turbulent BLs give rise to the logarithmic correction $\mathcal{L}(\text{Ta})$ in the scaling of the Nusselt number, which changes the scaling from $\text{Nu}_\omega\sim \text{Ta}^{1/2}$ to effectively $\text{Nu}_\omega\sim \text{Ta}^{1/2}\mathcal{L}(\text{Ta})\sim \text{Ta}^{0.40}$ \citep{VanGils2011,Huisman2012}. With \eref{eq:epsilon_global} one obtains the effective scaling of the global energy dissipation rate $\tilde{\epsilon}_{\text{global}}\sim \text{Ta}^{3/2} \mathcal{L}(\text{Ta})\sim \text{Ta}^{1.40}$. It is remarkable how our local measurements of the local energy dissipation rate reveal the very same scaling due to $\mathcal{L}(\text{Ta})$ as the global energy dissipation rate. In contrast, in RB flow it is shown that when the driving is in the order of $10^{8}<\text{Ra}<10^{11} $, i.e. far below the transition into the ultimate regime (BLs are still laminar), $\tilde{\epsilon}_{\text{bulk}}\sim \text{Ra}^{1.5}$ \citep{Shang2008,Ni2011}. Note, however, that in that regime the global energy dissipation rate $\tilde{\epsilon}_{\text{global}}$ is still determined by the BL contributions, $\tilde{\epsilon}_{\text{BL}}\gg\tilde{\epsilon}_{\text{bulk}}$ and $\tilde{\epsilon}_{\text{BL}}\approx \tilde{\epsilon}_{\text{global}}$. Our measurements are thus consistent with the prediction of \citet{Grossmann2011}, where even at such large Ta numbers, a rather intricate interaction between turbulent BLs and bulk flow prevails through the entire gap.

\begin{figure}
\begin{center}
\includegraphics[scale=0.4]{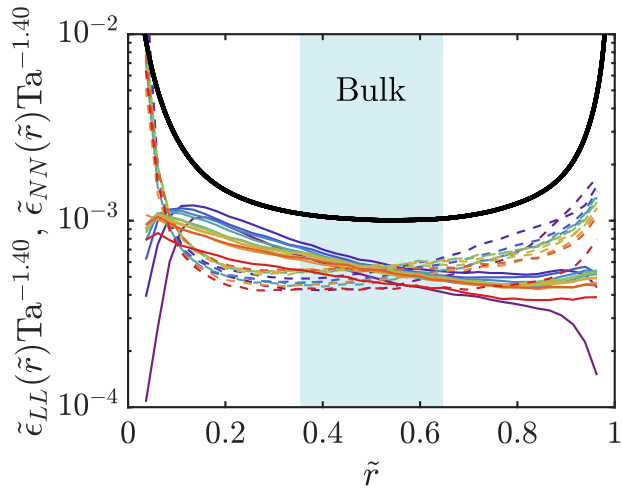}
\caption{(color online). Compensated dimensionless dissipation rate profiles calculated with both structure functions for different Ta: (dashed lines) longitudinal, (solid lines) transverse. The colors represent the variation in Ta as shown in \fref{fig:figure3}. In both figures, the bulk region is highlighted in blue.  The black solid line corresponds to the DNS data from  \citet{zhu2017} for $\text{Ta}=2.15\times 10^9$.}
\label{fig:figure6}
\end{center}
\end{figure}

In order to further show the quality of the scaling, we show in \fref{fig:figure6} the same $\epsilon$-profiles shown in \fref{fig:figure3} but now compensated with Ta$^{-1.40}$. For both the LL and NN direction, the dissipation rates for different Ta collapse throughout most of the gap, far away from the inner and outer cylinder. Within the bulk however, they are nearly constant and very close to the prefactors ($\approx 5\times 10^{-4}$) found from the scaling in \fref{fig:figure5}a. When looking at the compensated data from DNS, we notice that the prefactor is in that case twice as large as ours ($\approx 10^{-3}$). The reason is that the nature of both calculations is different: While the data from DNS is obtained from averaging the 3D velocity gradients over the entire volume, we rely on the scaling of the second order SFs (without intermittency corrections) to approximate the local energy dissipation rate in the bulk at the maximum peak in the compensated curves (see \S\ref{sec:sfs}).

In order to further characterize the turbulent scales in the flow, we calculate the Kolmogorov length scale in the bulk. Since there are two dissipation rates available, we define their corresponding Kolmogorov length scales as $\eta_{LL,\text{bulk}}=(\nu^3 / \epsilon_{LL,\text{bulk}})^{1/4}$ and $\eta_{NN,\text{bulk}}=(\nu^3 / \epsilon_{\text{NN,\text{bulk}}})^{1/4}$. Because $\tilde{\epsilon}_{\text{bulk}}\sim \text{Ta}^{1.40}$, the scaling of $\tilde{\eta}_{\text{bulk}}=\eta_{\text{bulk}} /d \sim \text{Ta}^{-0.35}$, which can be seen in \fref{fig:figure7}a.  Obviously, here we find a similar prefactor in both directions LL and NN too. The inset of the figure shows the corresponding compensated plot. For comparison, we include in the same figure the scaling from \citet{Lewis1999}. When comparing it with our data we notice some differences in magnitude. While we average in the bulk and make use of PIV to obtain the spatial dependence of the velocities directly, the data from \citet{Lewis1999} was measured at a single point ($\tilde{r}=0.5$) using Hot-wire anemometry and Taylor's frozen flow hypothesis.

When fitting data to a power law, confidence bounds for every coefficient in the regression can be obtained, given a certain confidence level. In this paper, we use the standard 95\% confidence for every fit, from which the uncertainty in the power law exponents (figures \ref{fig:figure5},\ref{fig:figure6}) were chosen as the middle point between the lower and upper bound of its corresponding confidence bound. This procedure is done for all the exponents reported throughout this paper.

\subsection{The turbulent intensity in the bulk}
\label{sec:eta_int}

\begin{figure}
\begin{center}
\includegraphics[scale=0.4]{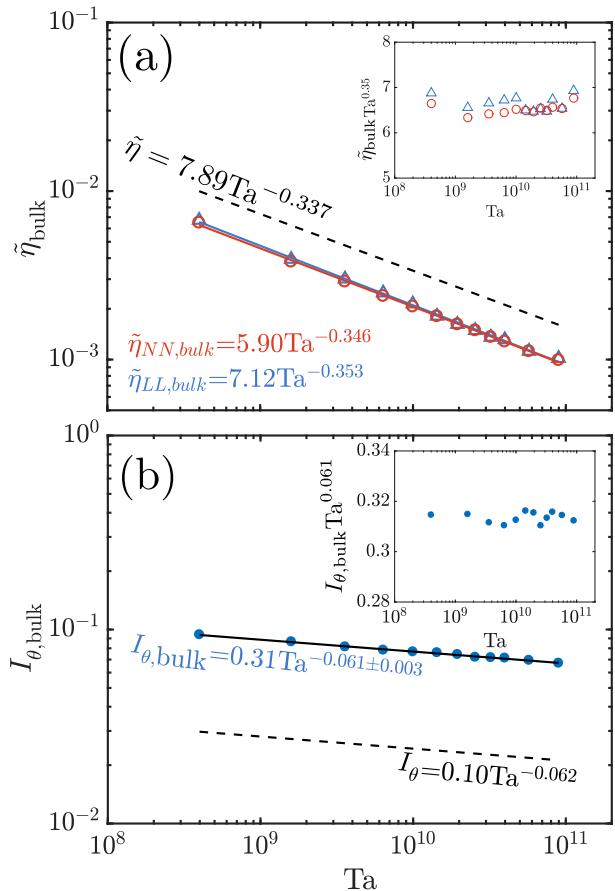}
\caption{(color online) (a) Dimensionless bulk-averaged Kolmogorov length scale: longitudinal (blue open triangles), transverse (red open circles). Local scaling at $\tilde{r}=0.5$ from \citet{Lewis1999} (black dashed line). The inset shows the compensated plots for the local quantities where the effective scaling of $\tilde{\eta}_{\text{bulk}}\sim\text{Ta}^{-0.35}$ is found to reproduce both directions. (b) Bulk-averaged azimuthal turbulent intensity. The data reveals an effective scaling of  $I_{\theta,\text{bulk}}\sim \text{Ta}^{-0.061}$. The (dashed black line) represents the local scaling $I_{\theta}=0.1$Ta$^{-0.062}$ at $\tilde{r}=0.5$ as it was obtained from \citet{Lewis1999}. The inset in (b) shows the corresponding compensated plot.}
\label{fig:figure7}
\end{center}
\end{figure}

The final step in the calculation of $\text{Re}_{\lambda,\text{bulk}}$ is to look at the azimuthal velocity fluctuations. Thus we average $\sigma_{\theta,t} (u_\theta(r,\theta,t))$ (see \eref{eq:sigma1}) from \fref{fig:figure2}a in the bulk and find a good description by the effective scaling law $(d/ \nu)\sigma_{\text{bulk}}(u_\theta) \approx 11.3 \times 10^{-2} \text{Ta} ^{0.44\pm0.01} $. In \fref{fig:figure7}b, we show the turbulence intensity $I_{\theta,{\text{bulk}}}=\langle \sigma_{\theta,t}(u_\theta) /\langle u_\theta\rangle_{\theta,t}  \rangle_{r_{\text{bulk}}}$ as a function of Ta. In this way, we are able to compare our data to the turbulence intensity scaling from \citet{Lewis1999}. We find that the effective scaling  $I_{\theta,\text{\text{bulk}}}\sim \text{Ta}^{-0.061\pm 0.003}$ reproduces our data well. In the inset of the same figure we show the compensated plot throughout the Ta range. Similarly as with the Kolmogorov length scale described in section \S\ref{sec:epsilon}, we include in the same figure the scaling of \citet{Lewis1999}. In this case, the exponent in our scaling is nearly identical to the one found by \cite{Lewis1999} with a slightly larger prefactor. We remind the reader once again that our average is done over the bulk region while the data of \citet{Lewis1999} is obtained at a single point at midgap.

\subsection{The scaling of the Taylor-Reynolds number $\text{Re}_{\lambda,\text{bulk}}$}
\label{sec:Relambda}

\begin{figure}
\begin{center}
\includegraphics[scale=0.4]{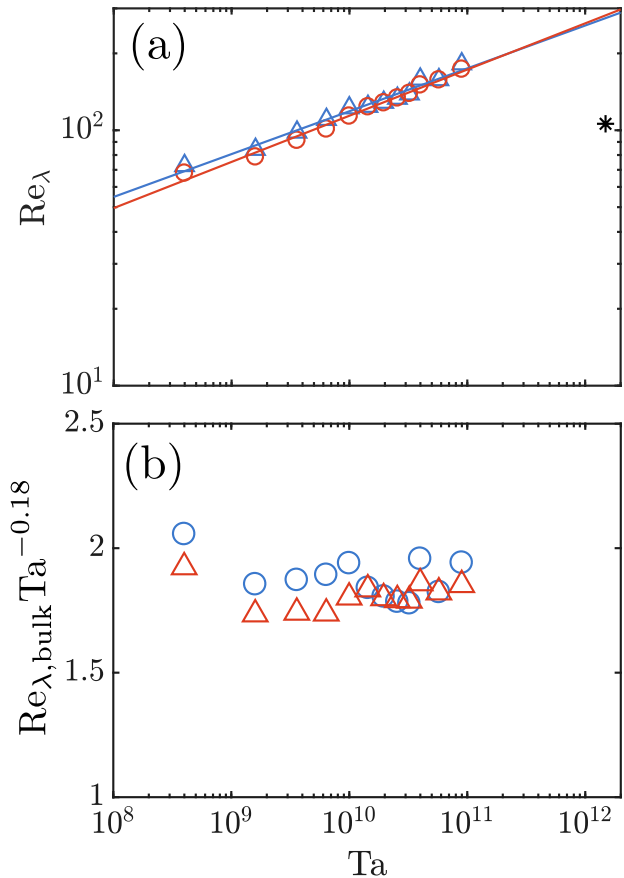}
\caption{(color online) (a) $\text{Re}_{\lambda,\text{bulk}}$ as a function of Ta. The blue open triangles (red open circles) show the calculation using $\epsilon_{LL,\text{bulk}}$ ($\epsilon_{NN,\text{bulk}}$). The black solid point is the calculation using the global energy dissipation rate from \citet{Huisman2013}. (b) Compensated plot of $\text{Re}_{\lambda,\text{bulk}}$ where an effective scaling of $\text{Re}_{\lambda,\text{bulk}}\sim \text{Ta}^{-0.18}$ is found to be in good agreement with both LL and NN directions.}
\label{fig:figure8}
\end{center}
\end{figure}

Finally, with both the local dissipation rate and the local velocity fluctuations in the bulk, we calculate the corresponding Taylor-Reynolds number as a function of Ta, using both $\epsilon_{LL,\text{bulk}}$, $\epsilon_{NN,\text{bulk}}$ and $\sigma_{\text{bulk}}(u_\theta)$. The results can be seen in \fref{fig:figure8}a where an effective scaling of $\text{Re}_{\lambda,\text{bulk}}\sim \text{Ta}^{0.18\pm 0.01}$ is found for both directions. The compensated plot in \fref{fig:figure8}b reveals the good quality of the scaling throughout the range of Ta. In order to highlight the difference between the different calculations, we also include the estimate of \citet{Huisman2013} for $\text{Ta}=1.49 \times 10^{12}$ ($\text{Re}_\lambda=106$). We emphasize that our calculation is based entirely on local quantities (fluctuations and dissipation rate) whilst the estimate of \citet{Huisman2013} is done using a single point in space, $\tilde{r}=0.5$, in combination with the global energy dissipation rate (\eref{eq:epsilon_global}). Our scaling predicts that the local Taylor-Reynolds number at that Ta is approximately $\text{Re}_{\lambda,\text{bulk}} \approx 217$, roughly twice the value estimated by \citet{Huisman2013} for the same Ta.

\section{Summary and conclusions}
\label{sec:conclusions}

To summarize,  we have measured local velocity fields using PIV in the ultimate regime of turbulence. We showed that both structure functions (longitudinal and transverse) yield similar energy dissipation rate profiles that intersect within the bulk, similarly as what is observed in Rayleigh-B\'{e}nard convection. When averaging these profiles within the bulk, this leads to an effective scaling of $\tilde{\epsilon}_{\text{bulk}} \sim \text{Ta}^{1.40\pm 0.04}$,  which is the same scaling as obtained for the global quantity $\tilde{\epsilon}$ measured from the torque scaling \citep{Ostilla-Monico2014a,Huisman2014}. This result reveals the dominant influence of the turbulent BLs over the entire gap. Future work will show whether this also holds for higher-order velocity structure functions, as it does hold in other turbulent wall-bounded flows \citep{sil15}.

Next, we showed that the Kolmogorov length scale scales as $\tilde{\eta}_{\text{bulk}} \sim \text{Ta}^{0.35\pm 0.01}$ and the azimuthal turbulent intensity scales as  $I_{\theta,\text{bulk}} \sim \text{Ta}^{-0.061\pm 0.003}$. In order to evaluate the turbulence level in the flow, we showed that with both local quantities at hand (dissipation rate and turbulent fluctuations), the bulk Taylor-Reynolds number scales as  $\text{Re}_{\lambda,\text{bulk}}\sim\text{Ta}^{0.18\pm 0.01}$. Our calculation can be generalized by inserting our result for the ratio between the local and global energy dissipation rate $\tilde{\epsilon}_{\text{bulk}}/ \tilde{\epsilon_{\text{global}}} =\alpha \approx 0.1$ back into \eref{eq:tare} and using \eref{eq:epsilon_global} to relate $\epsilon_{\text{global}}$ and $\text{Nu}_\omega$. The latter yields

\begin{equation}
\label{eq:tare_nusselt}
\text{Re}_{\lambda,\text{bulk}} (\text{Ta}) = \sqrt{1/\alpha} \left(\frac{\sqrt{15}  \sigma_{\text{TC}} d^2 }{\nu^2}\right) \frac{(\sigma_{\text{bulk}}(u_\theta))^2}{ \sqrt{\text{Ta} \text{Nu}_{\omega}}}.
\end{equation} 

\noindent Thus, given the local variance of the velocity fluctuations and the global Nusselt number, the response parameter $\text{Re}_{\lambda,\text{bulk}}(\text{Ta})$ can be calculated in the bulk flow ($\tilde{r} \in [0.35,0.65]$) for the case of pure inner cylinder rotation ($a=0$).  In order to extend the calculation to the case $a\approx a_{opt}\approx 0.36$, i.e. close to the rotation ratio for optimal $\text{Nu}_\omega$, where pronounced Taylor rolls exist \citep{Huisman2014,Rodolfo2014}, an extra averaging process in axial direction for both the velocity fluctuations and the dissipation rates would be needed.

\begin{acknowledgments}
We would like to thank B. Benschop, M. Bos and G.-W. Bruggert for their technical assistance. We acknowledge D. Bakhuis, R.A. Verschoof and R.C.A. van der Veen for stimulating discussions. We would also like to thank R. Ostilla-M\'onico and X. Zhu for making their DNS data available to us. This study was financially supported by the Fundamenteel Onderzoek der Materie (FOM). C. Sun acknowledges the financial support from Natural Science Foundation of China under Grant No. 11672156. 
\end{acknowledgments}

\bibliographystyle{jfm}
\bibliography{Relambda_JFM}

\begin{thebibliography}{35}
\expandafter\ifx\csname natexlab\endcsname\relax\def\natexlab#1{#1}\fi

\bibitem[Andereck {\em et~al.\/}(1986)Andereck, Liu \& Swinney]{Andereck1986}
{\sc Andereck, C.~D., Liu, S.~S. \& Swinney, H.~L.} 1986 Flow regimes in a
  circular \text{C}ouette system with independently rotating cylinders. {\em J.
  Fluid Mech.\/} {\bf 164}, 155--183.

\bibitem[Benzi {\em et~al.\/}(1993)Benzi, Ciliberto, Tripiccione, Baudet,
  Massaioli \& Succi]{Benzi1993}
{\sc Benzi, R., Ciliberto, S., Tripiccione, R., Baudet, C., Massaioli, F. \&
  Succi, S.} 1993 Extended self-similarity in turbulent flows. {\em Phys. Rev.
  E\/} {\bf 48}, R29--R32.

\bibitem[Blum {\em et~al.\/}(2010)Blum, Kunwar, Johnson \& Voth]{Blum2010}
{\sc Blum, D.~B., Kunwar, S.~B., Johnson, J. \& Voth, G.~A.} 2010 Effects of
  nonuniversal large scales on conditional structure functions in turbulence.
  {\em Phys. Fluids\/} {\bf 22}, 015107.

\bibitem[Chandrasekhar(1981)]{Chandrasekhar1961}
{\sc Chandrasekhar, S.} 1981 {\em Hydrodynamic and Hydromagnetic Stability\/}.
  Dover.

\bibitem[Chien {\em et~al.\/}(2013)Chien, Blum \& Voth]{Chien2013}
{\sc Chien, C.-C., Blum, D.~B. \& Voth, G.~A.} 2013 Effects of fluctuating
  energy input on the small scales in turbulence. {\em J. Fluid Mech.\/} {\bf
  737}, 527--551.

\bibitem[Eckhardt {\em et~al.\/}(2007)Eckhardt, Grossmann \&
  Lohse]{Eckhardt2007}
{\sc Eckhardt, B., Grossmann, S. \& Lohse, D.} 2007 {Torque scaling in
  turbulent \text{T}aylor-\text{C}ouette flow between independently rotating
  cylinders}. {\em J. Fluid Mech.\/} {\bf 581}, 221--250.

\bibitem[Fardin {\em et~al.\/}(2014)Fardin, Perge \& Taberlet]{Fardin2014}
{\sc Fardin, M.~A., Perge, C. \& Taberlet, N.} 2014 The hydrogen atom of fluid
  dynamics - introduction to the \text{T}aylor-\text{C}ouette flow for soft
  matter scientists. {\em Soft Matt.\/} {\bf 10}, 3523--3535.

\bibitem[Frisch(1995)]{Frisch1995}
{\sc Frisch, U.} 1995 {\em Turbulence: The Legacy of A. N. Kolmogorov\/}.
  Cambridge University Press.

\bibitem[van Gils {\em et~al.\/}(2011)van Gils, Huisman, Bruggert, Sun \&
  Lohse]{VanGils2011}
{\sc van Gils, D. P.~M., Huisman, S.~G., Bruggert, G.-W., Sun, C. \& Lohse, D.}
  2011 Torque scaling in turbulent \text{T}aylor-\text{C}ouette flow with co-
  and counterrotating cylinders. {\em Phys. Rev. Lett.\/} {\bf 106}, 024502.

\bibitem[van Gils {\em et~al.\/}(2012)van Gils, Huisman, Grossmann, Sun \&
  Lohse]{vangils2012}
{\sc van Gils, D. P.~M., Huisman, S.~G., Grossmann, S., Sun, C. \& Lohse, D.}
  2012 Optimal {T}aylor-{C}ouette turbulence. {\em J. Fluid Mech\/} {\bf 706},
  118--149.

\bibitem[Grossmann \& Lohse(2011)]{Grossmann2011}
{\sc Grossmann, S. \& Lohse, D.} 2011 Multiple scaling in the ultimate regime
  of thermal convection. {\em Phys. Fluids\/} {\bf 23}, 045108.

\bibitem[Grossmann {\em et~al.\/}(2016)Grossmann, Lohse \& Sun]{Grossmann2016}
{\sc Grossmann, S., Lohse, D. \& Sun, C.} 2016 High-\text{R}eynolds number
  \text{T}aylor-\text{C}ouette turbulence. {\em Annu. Rev. Fluid Mech.\/} {\bf
  48}, 53--80.

\bibitem[Huisman {\em et~al.\/}(2012)Huisman, van Gils, Grossmann, Sun \&
  Lohse]{Huisman2012}
{\sc Huisman, S.~G., van Gils, D. P.~M., Grossmann, S., Sun, C. \& Lohse, D.}
  2012 Ultimate turbulent \text{T}aylor-\text{C}ouette flow. {\em Phys. Rev.
  Lett.\/} {\bf 108}, 024501.

\bibitem[Huisman {\em et~al.\/}(2013)Huisman, Lohse \& Sun]{Huisman2013}
{\sc Huisman, S.~G., Lohse, D. \& Sun, C.} 2013 Statistics of turbulent
  fluctuations in counter-rotating \text{T}aylor-\text{C}ouette flows. {\em
  Phys. Rev. E\/} {\bf 88}, 063001.

\bibitem[Huisman {\em et~al.\/}(2015)Huisman, van~der Veen, Bruggert, Lohse \&
  Sun]{Huisman2015}
{\sc Huisman, S.~G., van~der Veen, R. C.~A., Bruggert, G.-W., Lohse, D. \& Sun,
  C.} 2015 The boiling \text{T}wente \text{T}aylor-\text{C}ouette ({BTTC})
  facility: Temperature controlled turbulent flow between independently
  rotating, coaxial cylinders. {\em Rev. Sci. Instrum.\/} {\bf 86}, 065108.

\bibitem[Huisman {\em et~al.\/}(2014)Huisman, van~der Veen, Sun \&
  Lohse]{Huisman2014}
{\sc Huisman, S.~G., van~der Veen, R. C.~A., Sun, C. \& Lohse, D.} 2014
  Multiple states in highly turbulent {Taylor-Couette} flow. {\em Nat.
  Commun.\/} {\bf 5}, 3820.

\bibitem[Lewis \& Swinney(1999)]{Lewis1999}
{\sc Lewis, G.~S. \& Swinney, H.~L.} 1999 Velocity structure functions,
  scaling, and transitions in high-\text{R}eynolds-number {Couette-Taylor}
  flow. {\em Phys. Rev. E\/} {\bf 59}, 5457--5467.

\bibitem[Mart\'{i}nez~Mercado {\em et~al.\/}(2012)Mart\'{i}nez~Mercado,
  Prakash, Tagawa, Sun \& Lohse]{martinez2012a}
{\sc Mart\'{i}nez~Mercado, J., Prakash, V.~N., Tagawa, Y., Sun, C. \& Lohse,
  D.} 2012 Lagrangian statistics of light particles in turbulence. {\em Phys.
  Fluids\/} {\bf 24}, 055106.

\bibitem[Ni {\em et~al.\/}(2011)Ni, Huang \& Xia]{Ni2011}
{\sc Ni, R., Huang, S.-D. \& Xia, K.-Q.} 2011 Local energy dissipation rate
  balances local heat flux in the center of turbulent thermal convection. {\em
  Phys. Rev. Lett.\/} {\bf 107}, 174503.

\bibitem[Ostilla-M{\'{o}}nico {\em et~al.\/}(2014)Ostilla-M{\'{o}}nico, van~der
  Poel, Verzicco, Grossmann \& Lohse]{Ostilla-Monico2014a}
{\sc Ostilla-M{\'{o}}nico, R., van~der Poel, E.~P., Verzicco, R., Grossmann, S.
  \& Lohse, D.} 2014 Boundary layer dynamics at the transition between the
  classical and the ultimate regime of {Taylor-Couette} flow. {\em Phys.
  Fluids\/} {\bf 26}, 015114.

\bibitem[Ostilla-M\'onico {\em et~al.\/}(2014)Ostilla-M\'onico, van~der Poel,
  Verzicco, Grossmann \& Lohse]{Rodolfo2014}
{\sc Ostilla-M\'onico, R., van~der Poel, E.~P., Verzicco, R., Grossmann, S. \&
  Lohse, D.} 2014 Exploring the phase diagram of fully turbulent
  {Taylor-Couette} flow. {\em J. Fluid Mech.\/} {\bf 761}, 1--26.

\bibitem[Paoletti \& Lathrop(2011)]{Paoletti2011}
{\sc Paoletti, M.~S. \& Lathrop, D.~P.} 2011 Angular momentum transport in
  turbulent flow between independently rotating cylinders. {\em Phys. Rev.
  Lett.\/} {\bf 106}, 024501.

\bibitem[Pope(2000)]{Pope2000}
{\sc Pope, S.~B.} 2000 {\em Turbulent Flows\/}. Cambridge University Press.

\bibitem[Shang {\em et~al.\/}(2008)Shang, Tong \& Xia]{Shang2008}
{\sc Shang, X.-D., Tong, P. \& Xia, K.-Q.} 2008 Scaling of the local convective
  heat flux in turbulent {Rayleigh-Bernard} convection. {\em Phys. Rev.
  Lett.\/} {\bf 100}, 244503.

\bibitem[She \& Leveque(1994)]{Z.She1994}
{\sc She, Z.-S. \& Leveque, E.} 1994 Universal scaling laws in fully developed
  turbulence. {\em Phys. Rev. Lett.\/} {\bf 72}, 336--339.

\bibitem[de~Silva {\em et~al.\/}(2015)de~Silva, Marusic, Woodcock \&
  Meneveau]{sil15}
{\sc de~Silva, C.~M., Marusic, I., Woodcock, J.~D. \& Meneveau, C.} 2015
  Scaling of second- and higher-order structure functions in turbulent boundary
  layers. {\em J. Fluid Mech.\/} {\bf 769}, 654--686.

\bibitem[Smith \& Townsend(1982)]{Smith1982}
{\sc Smith, G.~P. \& Townsend, A.~A.} 1982 Turbulent {Couette} flow between
  concentric cylinders at large \text{T}aylor numbers. {\em J. Fluid Mech.\/}
  {\bf 123}, 187--217.

\bibitem[Sreenivasan(1995)]{Sreenivasan1995}
{\sc Sreenivasan, K.~R.} 1995 On the universality of the {Kolmogorov} constant.
  {\em Phys. Fluids\/} {\bf 7}, 2778--2784.

\bibitem[Taylor(1923)]{GITaylor1923}
{\sc Taylor, G.~I.} 1923 Stability of a viscous liquid contained between two
  rotating cylinders. {\em Phil. Trans. R. Soc. A\/} {\bf 223}, 289--343.

\bibitem[van~der Veen {\em et~al.\/}(2016{\natexlab{{\em a\/}}})van~der Veen,
  Huisman, Dung, Tang, Sun \& Lohse]{veen2016a}
{\sc van~der Veen, R. C.~A., Huisman, S.~G., Dung, O.-Y., Tang, H.~L., Sun, C.
  \& Lohse, D.} 2016{\natexlab{{\em a\/}}} Exploring the phase space of
  multiple states in highly turbulent {Taylor-Couette} flow. {\em Phys. Rev.
  Fluids\/} {\bf 1}, 024401.

\bibitem[van~der Veen {\em et~al.\/}(2016{\natexlab{{\em b\/}}})van~der Veen,
  Huisman, Merbold, Harlander, Egbers, Lohse \& Sun]{vdveen2016}
{\sc van~der Veen, R. C.~A., Huisman, S.~G., Merbold, S., Harlander, U.,
  Egbers, C., Lohse, D. \& Sun, C.} 2016{\natexlab{{\em b\/}}}
  {T}aylor-{C}ouette turbulence at radius ratio ${\it\eta}=0.5$ :scaling, flow
  structures and plumes. {\em J. Fluid Mech\/} {\bf 799}, 334–351.

\bibitem[Voth {\em et~al.\/}(2002)Voth, La~Porta, Crawford, Alexander \&
  Bodenschatz]{Voth2002}
{\sc Voth, G.~A., La~Porta, A., Crawford, A.~M., Alexander, J. \& Bodenschatz,
  E.} 2002 Measurement of particle accelerations in fully developed turbulence.
  {\em J. Fluid Mech.\/} {\bf 469}, 121--160.

\bibitem[Zhou {\em et~al.\/}(2008)Zhou, Sun \& Xia]{zhou2008}
{\sc Zhou, Q., Sun, C. \& Xia, K.-Q.} 2008 Experimental investigation of
  homogeneity, isotropy, and circulation of the velocity field in
  buoyancy-driven turbulence. {\em J. Fluid Mech\/} {\bf 598}, 361–372.

\bibitem[Zhu {\em et~al.\/}(2017)Zhu, Verzicco \& Lohse]{zhu2017}
{\sc Zhu, X., Verzicco, R. \& Lohse, D.} 2017 Disentangling the origins of
  torque enhancement through wall roughness in {T}aylor–{C}ouette turbulence.
  {\em J. Fluid Mech\/} {\bf 812}, 279–293.

\bibitem[Zimmermann {\em et~al.\/}(2010)Zimmermann, Xu, Gasteuil, Bourgoin,
  Volk, Pinton \& Bodenschatz]{zimmermann2010}
{\sc Zimmermann, R., Xu, H., Gasteuil, Y., Bourgoin, M., Volk, R., Pinton,
  J.-F. \& Bodenschatz, E.} 2010 The lagrangian exploration module: An
  apparatus for the study of statistically homogeneous and isotropic
  turbulence. {\em Rev. Sci. Instrum.\/} {\bf 81}, 055112.

\end{thebibliography}

\end{document}